\documentclass[aps, notitlepage,nofootinbib, 11pt, a4paper]{revtex4-1}
\usepackage{graphicx}
\usepackage{float}
\usepackage{amsmath}
\usepackage{color}
\usepackage{tensor}
\usepackage{epstopdf}
\usepackage{xcolor}
\usepackage{wasysym}
\usepackage{hyperref}
\hypersetup{
	colorlinks,
	linkcolor={blue!50!black},
	citecolor={red!50!black},
	urlcolor={blue!80!black}
}

\usepackage{multibib}

\newcommand{\be}{\begin{equation}}
\newcommand{\ee}{\end{equation}}
\newcommand{\ba}{\begin{eqnarray}}
\newcommand{\ea}{\end{eqnarray}}

\newcommand{\beq}{\begin{equation}}
\newcommand{\eeq}{\end{equation}}
\newcommand{\beqa}{\begin{eqnarray}}
\newcommand{\eeqa}{\end{eqnarray}}
\newcommand{\nn}{\nonumber}

\begin{document}
\title{Gravitational Lensing by Black Holes in Einstein Quartic Gravity}

\pacs{04.50.Gh, 04.70.-s, 05.70.Ce}

\author{H. Khodabakhshi $^{1,2,3}$, Robert B. Mann$^{3,4}$
	\\
	\small$^1$Department of Physics, University of Tehran, Tehran, Iran\\ \small$^2$School of Particles and Accelerators, Institute for Research in Fundamental Sciences (IPM), Tehran, Iran\\\small$^3$Department of Physics and Astronomy, University of Waterloo, Waterloo, Ontario, Canada, N2L 3G1\\ \small$^4$Perimeter Institute, 31 Caroline St. N., Waterloo, Ontario, N2L 2Y5, Canada\\
}

\begin{center}
\begin{abstract}
	\hrulefill
	\section*{Abstract}
 We investigate gravitational lensing effects of spherically symmetric black holes in  Einstein Quartic Gravity (EQG).   
 Using an approximate analytic solution obtained by   continued fraction methods  we consider the  predictions of EQG for lensing effects by supermassive black holes at the center of our galaxy and others  in comparison with general relativity (GR). We  numerically compute both time delays and   angular positions of images and find that they can  deviate from GR by as much as milliarcseconds, suggesting that observational tests of EQG are feasible in the near future. We discuss the challenges of distinguishing the predictions of EQG from those of Einstein Cubic Gravity. 	
 
	\hrulefill
\end{abstract}
\end{center}

\maketitle

\section{Introduction}

Originally the phenomenon of gravitational lensing (GL), namely the bending of light \cite{eddington}, 
 was the most significant demonstration of the validity of General Relativity  (GR) \cite{Einstein}. It has since become a fruitful and primary tool for studying some 
 of the most important aspects of cosmology and astrophysics, such as the distribution of dark matter in  galaxy clusters \cite{Falco}. The phenomenon has been
 studied in both weak field and strong field regimes \cite{Dyson}. For   strong gravitational fields an infinite number of images (called relativistic images)
 on each side of the optical axis of a Schwarzschild black hole have been found  \cite{Darwin59,Darwin61,Ellis}. A calculation of time delay between the outermost two relativistic images has been useful in obtaining the mass of the black hole with high precision. Furthermore, a given mass and   angular separation between relativistic images can be used to calculate the distance to the black hole \cite{Keeton,Virbhadra}.

Strong gravitational fields are significantly modified by higher curvature corrections.   The most well known such corrections are given by the Lovelock class of theories~\cite{Lovelock:1971yv}.  This class has a number of noteworthy features of interest, including having second order differential equations and 
a particle spectrum that is the same as Einstein gravity. However from a phenomenological perspective they have the disadvantage that they are trivial in
4 spacetime dimensions.   Recently two newer classes of higher-curvature gravity have been discovered.  One is
 Quasi-topological gravity \cite{Myers:2010ru, Oliva:2010eb, Cisterna:2017umf}, whose formulation is
 also in more than 4 dimensions. Another is \textit{Generalized Quasi-Topological Gravity}  (GQTG)~\cite{Pablo1,Hennigar:2017ego, Ahmed:2017jod}, which 
are constructed by requiring that there is a single independent field equation for only one metric function under the restriction of spherical symmetry.  They
 have  attracted interest because  such theories  have the same graviton spectrum as general relativity on constant curvature backgrounds and are non-trivial in 4 spacetime dimensions. As such they provide a new set of phenomenological competitors to general relativity in strong-field regimes, whose parameters can be constrained by observation.

Here we investigate   GL of black holes in Einstein Quartic Gravity (EQG), the next simplest GQTG after {\it Einsteinian cubic gravity} (ECG)~\cite{Pablo1}. 
Although  the systematic construction of actions that are $n$-th order in curvature  from lower order ones via
recursive formulas have been  obtained that  allow for construction of any GQTG
 \cite{Bueno:2019ycr}, EQGs  have the highest degree of curvature  possible that allows for an analytic solution of the near horizon equations for the temperature and mass in terms of the horizon radius $r_+$.  As such they are of particular interest and merit further study.
We also note that the  EQG theory we study does not meet the general criteria used to construct an alternative class of ``Einsteinian" higher-curvature theories \cite{Bueno:2016ypa} (though its cubic counterpart \cite{Hennigar:2017ego} satisfies these criteria), apart from a   particular quartic GQTG. We therefore expect that some combination of  the EQG  invariants we consider (perhaps with some possibly trivial densities) could satisfy these other criteria.    

The Lagrangian of EQG is
\begin{equation}
\label{lagden}
{\cal L} = \frac{1}{16\pi}\left[R - \sum_{i=1}^{6} \hat{\lambda}_{(i)} \mathcal{S}_{4}^{(i)}\right],
\end{equation}
where $R$ is the usual Ricci scalar and $\mathcal{S}_{4}^{(i)}$ are called \textit{quasi-topological Lagrangian densities}  \cite{Ahmed:2017jod}.
 Clearly there are six such quartic curvature combinations that are nontrivial in (3+1) dimensions, leading to the introduction of six dimension-independent new coupling constants. Under the imposition of spherical symmetry 
 the field equations  differ by terms that vanish for a static spherically symmetric (SSS) metric, leading to a
  degeneracy that yields one new effective coupling constant that is a linear combination of the six couplings.  We shall henceforth only consider this case.

Although the technical challenges in solving GTQG equations are formidable, even if spherical symmetry is imposed,  approximate analytic
 solutions to the field equations of ECG \cite{OURS,POSHTEH,Adair:2020vso} and EQG  \cite{Khodabakhshi:2020hny} have been obtained using continued fraction methods~\cite{Rezzolla:2014mua, Kokkotas:2017zwt, Kokkotas:2017ymc, Konoplya:2016jvv}. This type of solution provides an excellent approximation 
to the actual solution  everywhere outside the horizon provided the continued fraction is taken to sufficiently high order.

 GL effects have been investigated analytically in the strong field limit approximation~\cite{Bozza2002,Bozza2001} 
 for many different black holes in GR and alternative theories~\cite{Torres,Bozza2003,Bhadra,Whisker,Eiroa,Jing2009,Wei,Cai,Jing2017}, but have also  been criticized for their accuracy~\cite{Virbhadra}. In what follows we shall employ the continued fraction solution \cite{Khodabakhshi:2020hny} 
 adapting methods developed for Schwarzschild black holes~\cite{Ellis} to investigate GL by black holes in EQG.

We find that   the difference between the angular positions of primary and secondary images in EQG and GR could be as large as milliarcseconds for
values of the EQG parameter consistent with other observations. Furthermore, the predicted values of time delay between these images in GR and EQG could be as large as seconds for a lots of number of angular source position. Our results indicate that  observational tests of EQG are no less feasible than
for  ECG  \cite{POSHTEH}. We also  compare the predictions of EQG with those of  ECG and show that these two cases are marginally distinguishable at best.

Our paper is organized as follows: In section 2 we give a review of the continued fraction method to obtain the approximate analytic asymptotically flat, static and spherically symmetric vacuum solution (SSS) to EQG. In Section 3 we consider the Lagrangian of massless particle to calculate equations needed to study the GL effects such as the relation for the bending angle, time delay and magnification of images. In next section, using these equations we investigate GL of SMBHs, for  Sgr A* and those at the centers of thirteen other galaxies and in last section we explain our conclusion and remarks. Our calculations are in units where $G = c= 1$.

\section{Black Hole Solution in Einstein Quartic Gravity}

We review here the continued fraction method for obtaining the metric function of EQG under the ansatz of spherical symmetry \cite{Khodabakhshi:2020hny}.  Consider an asymptotically flat, static and spherically symmetric vacuum black hole whose metric is of the form
\begin{equation}\label{eqn:le}
ds^2 = -f(r) dt^2 +\frac{dr^2}{f(r)} + r^2\left(d\theta^2 + \sin^2\theta d\phi^2 \right),
\end{equation}
in which we have $\lim_{r\to\infty}f(r) = 1$. Substituting this metric into the Lagrangian~\eqref{lagden}, the field equation for EQG can be written, after performing
an integration,  as
\begin{align}\label{eqn:feq}
r(1-f) - \frac{24}{5} K \bigg[\frac{1}{r^2} ff'f''(f-1-\frac{1}{2}rf')+ \frac{1}{8r} f'^4 + \frac{1}{6 r^2}f'^3(f+2)+\frac{1}{r^3} f f'^2 (1-f)\bigg] = 2 M,
\end{align} 
where $M$ is a constant of integration  and the prime denotes differentiation with respect to $r$. The constant $K$ is a linear combination of the six EQG coupling constants  
\begin{align}
{\lambda}_{(1)} &= -\frac{6}{5} \hat{\lambda}_{(1)}\,, \; \; \; {\lambda}_{(2)} = -3\hat{\lambda}_{(2)}\,, \; \; \; {\lambda}_{(3)} = -\frac{12}{5} \hat{\lambda}_{(3)}\,, \; \; \; {\lambda}_{(4)} = -\frac{24}{5} \hat{\lambda}_{(4)}\,, \; \; \; \nonumber \\{\lambda}_{(5)} &= -\frac{24}{5} \hat{\lambda}_{(5)}\,, \; \; \; {\lambda}_{(6)} = -\frac{96}{5} \hat{\lambda}_{(6)}\,.
\end{align}
and is
\begin{align}\label{EQGK}
K\equiv -\frac{5}{6} \left(\sum_{i=1}^{6} \lambda_{(i)} \right) \, ,
\end{align}
because each term $\mathcal{S}_{4}^{(i)}$ has the same contribution to the field equation under spherical symmetry~\cite{Ahmed:2017jod}.
The quantity $M$ in the field equation is the ADM mass of the black 
hole~\cite{Bueno:2016lrh, Hennigar:2017ego}. We should assume that $K > 0$ if we consider an asymptotically flat  solution.

To obtain the continued fraction solution, we begin with the near horizon series expansion of the metric function  
\begin{align}\label{eqn:nh_expand} 
f_{\rm nh}(r) = 4 \pi T (r-r_+) + \sum_{n=2}^{n = \infty} a_n (r-r_+)^n \, ,
\end{align} 
where $T = f'(r_+)/4\pi$ is the Hawking temperature. Upon inserting this ansatz into the field equations~\eqref{eqn:feq}, we   obtain
\begin{align}\label{eqn:mass_temp}
T &=\frac{1}{4 \pi r_+} \left[ \frac{1}{2} (\xi - \sqrt{\tau})-2 \right]\,,
\nonumber\\
M &=\frac{r_+}{2} \Bigl( -\frac{2048 K }{5 r_+^6} - 20 \Bigr) + \sqrt{\tau} \Bigl( -\frac{32 K}{r_+^5} + \frac{2 (25K)^\frac{1}{3}}{5r_+} -\frac{3r_+}{4} - \frac{8 K}{(25K)^\frac{1}{3} r_+^3} \Bigr) 
\nonumber\\
&+\frac{1}{\sqrt{\tau}} \Bigl( -\frac{1536 K}{5 r_+^5} - 24 r_+ \Bigr) + \xi \Bigl(  \frac{128K}{5 r_+^5} -\frac{8K}{(25K)^\frac{1}{3} r_+^3} + \frac{2(25K)^\frac{1}{3} }{5r_+} +\frac{3r_+}{4} \Bigr) 
\nonumber\\ 
&+\frac{\xi}{\sqrt{\tau}} \Bigl( \frac{128K}{5 r_+^5} + 2r_+ \Bigr) + (\xi \sqrt{\tau}) \frac{24K}{5 r_+^5} \,,
\end{align}
in which we have
\begin{align}
\tau \equiv 16 - \frac{20}{(25K)^\frac{1}{3}} r_+^2 + \frac{(25K)^\frac{1}{3}}{K} r_+^4 \qquad 
\xi \equiv \sqrt{ 48 + \frac{128}{\sqrt{\tau}} + \frac{10}{K \sqrt{\tau}} \, r_+^6 - \tau} \,,
\end{align}
determining temperature and mass in terms of $r_+$ and $K$.
Note that all $a_n$ for $n > 2$ can be determined from the field equation in terms of $T(r_+,K)$, $M(r_+,K)$, $r_+$, and $a_2$; the constant $a_2$ 
cannot be so determined.
 
Likewise we can write the asymptotic solution to \eqref{eqn:feq} as~\cite{Hennigar:2017ego,Khodabakhshi:2020hny}
\begin{equation}
f(r) \approx 1 - \frac{2 M}{r}- \frac{864}{5} \frac{K M^3}{r^9} + \frac{1552}{5} \frac{K }{r^{10}} + O \left( \frac{K^2 \, M^5} {r^{17}} \right)\,.
\end{equation}
We can match this solution to the near horizon approximation by  numerically solving the equations of motion in the intermediate regime. We do so by picking a value for $a_2$ for given values of $M$ and $K$ and use these in the near horizon expansion to obtain the initial data
\begin{align}\label{eqn:nh_data}
f(r_+ + \epsilon) &= 4 \pi T \epsilon + a_2 \epsilon^2 \, ,
\nn\\
f'(r_+ + \epsilon) &= 4 \pi T  + 2 a_2 \epsilon \, ,
\end{align}
in which $\epsilon$ is some small, positive quantity.  We find that  \cite{Khodabakhshi:2020hny}
\begin{equation}
\label{eqn:approx_a2} 
a^{*}_2\left(x=K/M^6\right)=-\frac{1}{M^2}\frac{1 + 2.23817 x + 0.0322907 x^2}{4 + 15.0556 x + 6.70964 x^2}\,,
\end{equation}
is the unique value of $a_2$ for which the numerical solution  agrees with the asymptotic expansion at a sufficiently large value of $r$, where   
the expression \eqref{eqn:approx_a2} is accurate to at least three decimal places   in the interval $K/M^6 \in [0, 5]$.

To obtain an approximate  continued fraction solution we  compactify the space-time interval outside of the horizon using the coordinate $x=1-r_+/r$, and then write
\begin{equation}
\label{eqn:cfrac_ansatz} 
f(x) = x \left[1 - \varepsilon(1-x) + (b_0 - \varepsilon)(1-x)^2 + \tilde{B}(x)(1-x)^3 \right],
\end{equation}
where
\begin{equation}
\tilde{B}(x) = \cfrac{b_1}{1+\cfrac{b_2 x}{1+\cfrac{b_3 x}{1+\cdots}}} \,.
\end{equation}
Inserting the ansatz~\eqref{eqn:cfrac_ansatz} into the field equation~\eqref{eqn:feq} yields
\begin{align}
\varepsilon &= \frac{2 M}{r_+} - 1,  \qquad b_0 = 0 \,,
\end{align}
and by expanding~\eqref{eqn:cfrac_ansatz} near the horizon ($x=0$), successive terms in the expansion provide expressions for all coefficients  in terms of $T$, $M$, $r_+$ and one free parameter, $b_2$, given by 
\begin{align}\label{eqn:frac_b2}
b_1 = 4 \pi r_+ T + \frac{4 M}{r_+} - 3, \hspace{1cm} b_2 = - \frac{r_+^3 a_2 + 16 \pi r_+^2 T + 6(M-r_+)}{4 \pi r_+^2 T + 4 M - 3 r_+} \, .
\end{align}
where we see that $b_2$ is dependent to the coefficient $a_2$ appearing in the near horizon expansion~\eqref{eqn:nh_expand}. We obtain
the relevant value of $b_2$ from $a_2^\star$ (as determined  numerically). While   numerical integration of the field equations is quite sensitive to the precision of $a_2^\star$, the continued fraction is much less so, and a good approximation is obtained even with just a few significant digits.

\section{Black hole lensing}

In this section we review briefly some basic equations needed to study GL by black holes \cite{POSHTEH}. The Lagrangian can be written as 
\be\label{Lagnull}
2\mathcal{L}=g_{\mu\nu}\dot{x}^\mu\dot{x}^\nu = -f \dot{t}^2 + \frac{\dot{r}^2}{f} + r^2 \dot{\phi}^2
\ee
using (\ref{eqn:le}), assuming for simplicity that the observer, black hole and the source  are on the equatorial plane $\vartheta=\pi/2$. For null geodesics   we   obtain  
\begin{align}\label{eqn:lagf1}
\frac{1}{fr^2}\left(\frac{dr}{d\phi}\right)^2=\dfrac{r^2}{f}\frac{E^2}{L_z^2}-1
\end{align}
where
\begin{align}\label{eqn:lagf2}
E &=- \frac{\partial \mathcal{L}}{\partial \dot{t}} = f\dot{t} \qquad  L_z =- \frac{\partial \mathcal{L}}{\partial \dot{\phi}} = - r^2\dot{\phi}  
\end{align}
are constants of the motion, with the overdot indicating a proper time derivative. 

\begin{figure}
	\centering
	\includegraphics[width=0.5\textwidth]{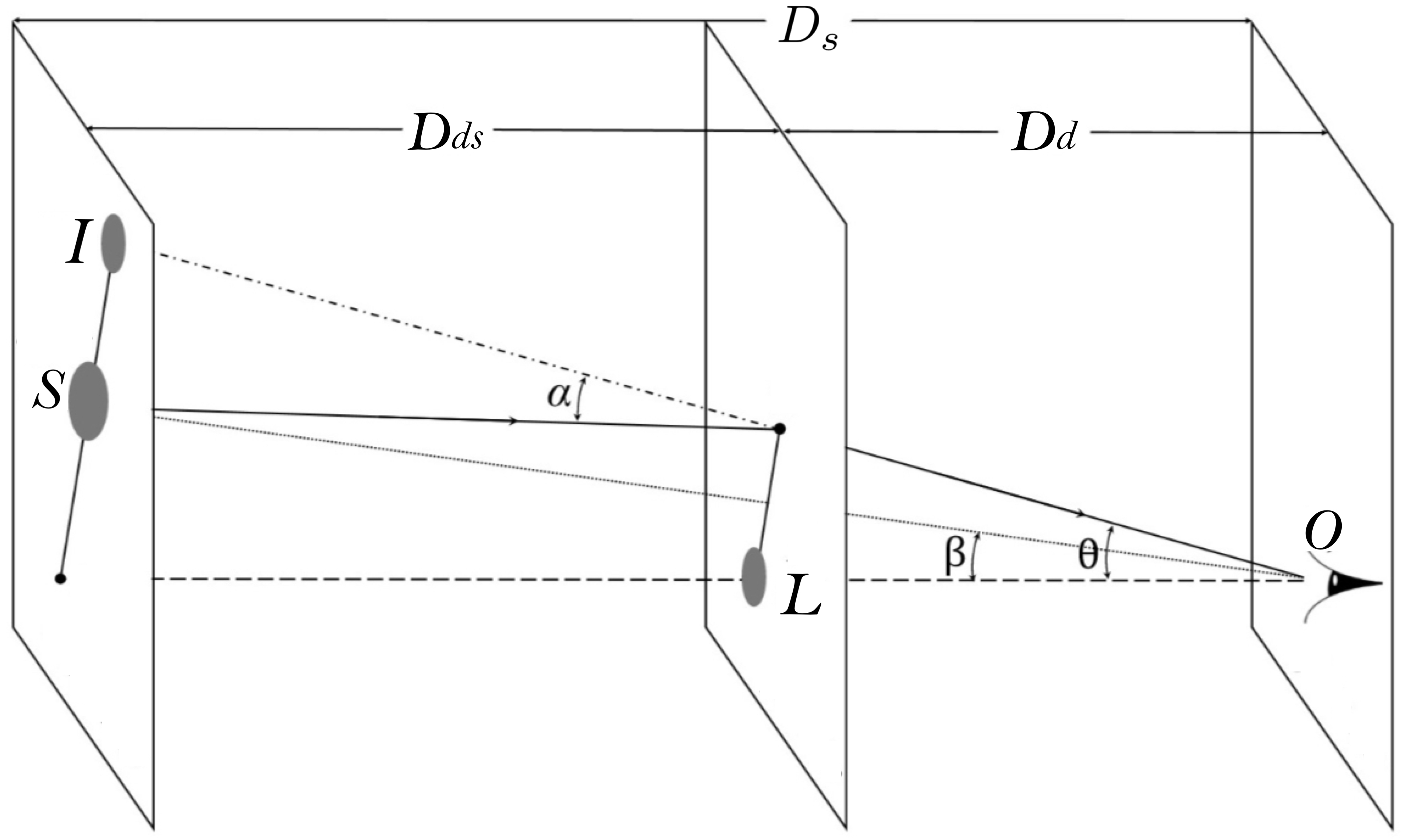}
	\caption{{\bf The lens diagram}: When a light ray passes a black hole it is deflected by an angle $\hat{\alpha}$, with rays passing closer to the black hole having a larger deflection angle. If $\hat{\alpha}>2\pi$, the corresponding light ray winds the black hole at least once, before reaching the observer -- these rays  make the relativistic images. In this figure $S$, $I$, $O$, and $L$ are the source, image, observer, and   lens (which is a black hole in our study), respectively.   $\beta$ is the angular position of the source w.r.t. the line of sight to the black hole and $\theta$ is the angular position of the image. $D_d$ and $D_{ds}$ demonstrate the distance from lens to observer and from lens to the source, respectively.}
	\label{fig:lens_diag}
\end{figure}

Since  $dr/d\phi=0$ at the radius of closest approach $r=r_0$, we have $E^2/L_z^2=f_0/r_0^2$ from \eqref{eqn:lagf1}, where $f_0 = f(r_0)$. Hence 
\begin{align}\label{eqn:dphidr}
\frac{d\phi}{dr}=\frac{1}{r\sqrt{\left(\frac{r}{r_0}\right)^2f_0-f}}.
\end{align}
A useful schematic diagram of the LG effect is exhibited in Fig.~\ref{fig:lens_diag}. $D_d$ and $D_{ds}$ demonstrate the distance of the lens (L) from the observer (O) and the source (S) respectively. By assuming $D_d, D_{ds}\gg r_0$, we can obtain the deflection angle ~\cite{weinberg1972}
\begin{align}\label{eqn:alpha_hat}
\hat{\alpha}(r_0)=2\int_{r_0}^{\infty}\frac{dr}{r\sqrt{\left(\frac{r}{r_0}\right)^2f_0-f}}-\pi.
\end{align}

Furthermore,  since $dr/dt=0$ at $r=r_0$ from \eqref{Lagnull} we can write
\begin{align}\label{eqn:dtdr}
\frac{dt}{dr}=\frac{1}{f\sqrt{1-\left(\frac{r_0}{r}\right)^2\frac{f}{f_0}}}.
\end{align}
The time delay is the difference between the time for the photons to travel the physical path from the source to the observer and the time it takes to reach the observer when there is no black hole in between them (i.e. in flat spacetime). Using   (\ref{eqn:dtdr})   the time delay of an image is
\begin{align}\label{eqn:time_delay}
\tau(r_0)=\left[\int_{r_0}^{r_s}dr+\int_{r_0}^{r_o}dr\right]\frac{1}{f\sqrt{1-\left(\frac{r_0}{r}\right)^2\frac{f}{f_0}}}-D_s\sec\beta,
\end{align}
where $D_s=D_d+D_{ds}$ is the distance from observer to the source, $r_s=\sqrt{D_{ds}^2+D_s^2\tan^2\beta}$, and $r_o=D_d$, with $\beta$ the angular position of the source.

The image angular position, $\theta$, obeys~\cite{Ellis}
\begin{align}\label{eqn:lens_eq}
\tan\beta=\tan\theta-\mathcal{D}\left[\tan\theta+\tan(\hat{\alpha}-\theta)\right],
\end{align}
where $\mathcal{D}=D_{ds}/D_s$. The impact parameter and the image magnification are given by~\cite{Virbhadra98}
\begin{align}\label{eqn:impact}
J=\frac{r_0}{\sqrt{f_0}}=D_d\sin\theta,
\end{align}
and
\begin{align}\label{eqn:maggi}
\mu=\left(\frac{\sin\beta}{\sin\theta}\frac{d\beta}{d\theta}\right)^{-1}
\end{align}
respectively, where  $\theta$  is  the deflection angle. 

We are interested in the rate of change of the deflection angle $\hat{\alpha}$ in \eqref{eqn:alpha_hat} with respect to $r_0$. This is somewhat subtle to compute, but
a series of manipulations \cite{POSHTEH} eventually yields
\begin{align}\label{eqn:alpha_r}
\frac{d\hat{\alpha}(r_0)}{dr_0}=-2\int_{r_0}^{\infty}\frac{dr}{\sqrt{\mathcal{F}}}\frac{\partial \tilde{\mathcal{F}}}{\partial r},
\end{align}
with
\begin{align}
\tilde{\mathcal{F}}=\frac{1}{r}\frac{\partial\mathcal{F}}{\partial r_0}\frac{\partial r}{\partial\mathcal{F}}.
\end{align}
where $\mathcal{F}=\left(\frac{r}{r_0}\right)^2f_0-f(r)$. In the following sections we will use these results to investigate  GL effects for  black holes in GR and EQG.

\section{Lensing by supermassive black holes}

Now we have the necessary tools to investigate lensing effects due to  supermassive black holes (SMBHs) at the center of the Milky Way and thirteen other galaxies.  Our main aim is to compare the lensing predictions of GR with those of EQG. To do so we numerically solve equations (\ref{eqn:alpha_hat}), (\ref{eqn:lens_eq}), (\ref{eqn:maggi}), and (\ref{eqn:time_delay}),  to respectively find their deflection angles, angular positions of their images, their magnifications, and their time delays. Although lensing due to Sgr A* in GR has been extensively investigated numerically \cite{Ellis,Keeton,Virbhadra,POSHTEH}, we recalculate the GR results with greater precision 
for the mass of Sgr A* $M=5.94\times 10^9 \, {\rm m}$, which is at a distance $D=2.43\times 10^{20} \, {\rm m}$ from Earth~\cite{mnd}.

We have previously shown \cite{Khodabakhshi:2020hny} that EQG passes all the Solar System tests to date if the coupling constant of EQG not to be larger than $K = 8.98 \times 10^{38} M_{\astrosun}^6$. We shall assume the largest possible value of $K$ to show that EQG lensing effects can differ significantly from the GR predictions. Furthermore, we compare   EQG lensing effects with those of ECG for the largest possible values of their respective coupling constants.

Using (\ref{eqn:alpha_hat}) and (\ref{eqn:lens_eq}) we compute the bending angle $\hat{\alpha}$ and angular image position $\theta$ for primary and secondary images, which are the respective images on the same and opposite sides of the source, taking ${\cal D} = D_{ds}/D_d  =0.5$; the means the lens-source distance is the same as the lens-observer distance, appropriate for    Sgr A*.
We compile the results in Table \ref{tab:Tablei} for both GR and EQG, with the coupling constant $K/M_{Sgr A*}^{6}\approx 2.21\times 10^{-1}$. We that in general the EQG results for the deflection angle and image angular positions ($\theta_p$ or $|\theta_s|$) are smaller than their corresponding values in GR. 
\begingroup
\begin{table*}
	\caption{
		{\bf Image positions and deflection angles of primary and secondary images due to lensing by Sgr A* with ${\cal D}=0.5$}: GR and EQG predictions for angular positions $\theta$ and bending angles $\hat{\alpha}$ are given for different values of angular source position $\beta$. {\bf (a)} $p$ and $s$ refer to primary and secondary images, respectively. {\bf (b)} All angles are in {\em arcseconds}. {\bf (c)} We have used $M_{Sgr A*}=5.94\times 10^9 \, {\rm m}$, $D_d=2.43\times 10^{20} \, {\rm m}$, and $K/M_{Sgr A*}^{6}\approx 2.21\times 10^{-1}$.
	}\label{tab:Tablei}
	\begin{tabular}{l cccc cccc}
		\hline \hline
		\multicolumn{1}{c}{$\beta$}& 
		\multicolumn{4}{c}{General relativity}& 
		\multicolumn{4}{c}{Einsteinian Quartic Gravity}\\
		&$\theta_{p,{\rm GR}} $&$\hat{\alpha}_{p,{\rm GR}}$&$\theta_{s,{\rm GR}}$&$\hat{\alpha}_{s,{\rm GR}}$&$\theta_{p,{\rm EQG}}$&$\hat{\alpha}_{p,{\rm EQG}}$&$\theta_{s,{\rm EQG}}$&$\hat{\alpha}_{s,{\rm EQG}}$\\
		\hline
		$0	 	 $&$  1.44324     $&$	2.88648    		$&$  -1.44324  $&$    	  2.88648     $&$  1.44291     $&$	  2.88620   	  $&$ -1.44291     $&$    2.88568     	   $\\
		$10^{-3} $&$  1.44374	  $&$   2.88548    		$&$  -1.44274  $&$	   	  2.88748     $&$  1.44341	   $&$    2.88241   	  $&$ -1.44241	   $&$	   2.88782        $\\
		$10^{-2} $&$  1.44825	  $&$   2.87650    		$&$  -1.43825  $&$		  2.89650     $&$  1.44792	   $&$    2.87608   	  $&$ -1.43792     $&$	   2.89523        $\\
		$10^{-1} $&$  1.49411	  $&$   2.78821    		$&$  -1.39411  $&$		  2.98821     $&$  1.49376	   $&$    2.78467   	  $&$ -1.39379     $&$	   2.99321      $\\
		$1	 	 $&$  2.02740	  $&$   2.05479    		$&$  -1.02740  $&$    	  4.05480     $&$  2.02691	   $&$    2.05382   	  $&$ -1.02719     $&$	   4.06073          $\\
		$2	 	 $&$  2.75583	  $&$   1.51166    		$&$  -0.755838 $&$    	  5.51167     $&$  2.75521	   $&$    1.51105   	  $&$ -0.75570    $&$    5.51005         $\\
		$3	 	 $&$  3.58157	  $&$   1.16314    		$&$  -0.581575 $&$		  7.16322     $&$  3.58087	   $&$    1.16225   	  $&$ -0.581473    $&$	   7.16069         $\\
		$4	 	 $&$  4.46636	  $&$   0.932720   		$&$  -0.466372 $&$    	  8.93274     $&$  4.46571	   $&$    0.92783  	  $&$ -0.46627	   $&$    8.93275        $\\
		\hline \hline
	\end{tabular}
\end{table*}
\endgroup

We have previously shown \cite{Khodabakhshi:2020hny} that the shadow of Sgr A* is enlarged in EQG by an amount less than 10 nanoarcseconds relative to GR for $K/M_{Sgr A*}^{6}\approx 2.21\times 10^{-1}$.
This occurs because the size of the shadow of Sgr A* is of order of $10^{-5}$ arcseconds  whether or not its gravitational field is governed by GR or EQG, and is far lower than the resolution of today's observational facilities such as Event Horizon Telescope \cite{eht,Akiyama}.
 
\begin{figure}
	\centering
	\includegraphics[width=1\textwidth]{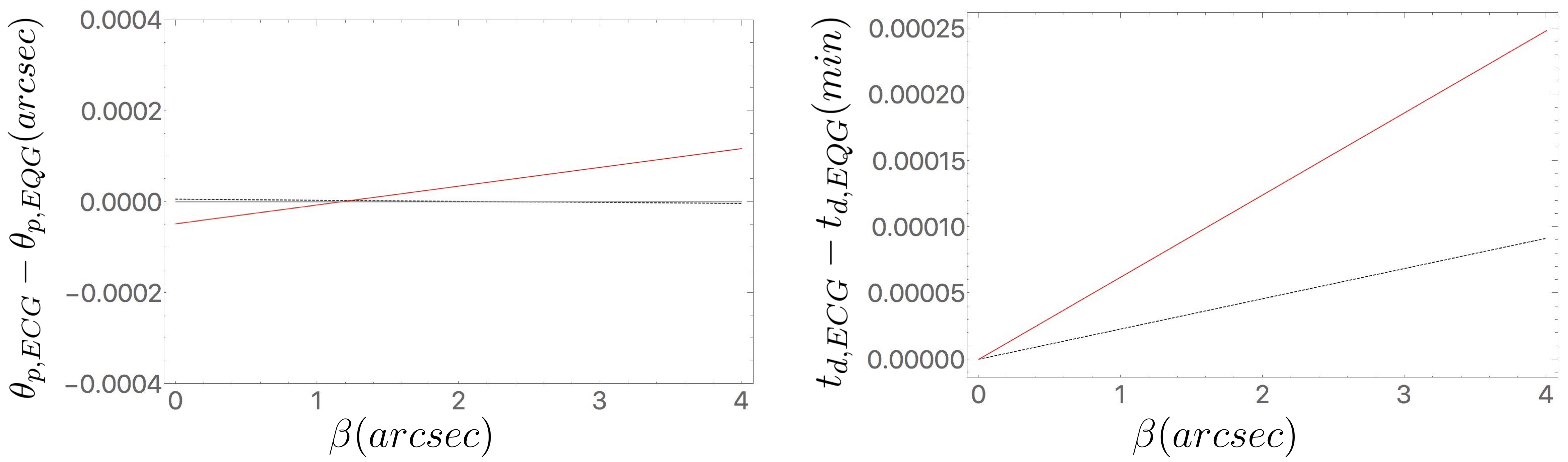}
	\caption{{\bf Deviation of primary image angular position and differential time delay in EQG from ECG for Sgr A*}:  The red line is for the case ${\cal D}=0.5$ and the black dotted line is for ${\cal D}=0.05$. \textbf{Left:} The deviation of EQG results for angular positions of primary images from that of ECG is larger for the sources further away from lens and is quite small for $\beta \approx 1$. \textbf{Right:} The deviation of EQG results for the differential time delay $t_d=\tau_s-\tau_p$ from that of ECG is larger for the sources further away from lens and increases by increasing $\beta$.
	}
	\label{fig:diff_thetaeq}
\end{figure}

However the source positions and the angular positions of primary/secondary images in GR or EQG are of the order of arcseconds, and so 
the difference between these angular positions (with the same value of $K$)  could be on the order of miliarcseconds. This is comparable to the ECG results
\cite{POSHTEH} and so differences between GR an EQG are potentially distinguishable with near-future observations.  However the differences between ECG and EQG are not easily distinguishable using SgrA*, as we shall see when we compare ECG and EQG in Fig.~\ref{fig:diff_thetaeq}. 

In Table \ref{tab:Tableii} we present our result for the magnification $\mu$ of the primary and secondary images of Table \ref{tab:Tablei}, computed using  (\ref{eqn:maggi})  and (\ref{eqn:alpha_r}) and the time delay $\tau$ of the primary images from (\ref{eqn:time_delay}). Since the difference $t_d=\tau_s-\tau_p$ between the time delay of the secondary and the primary images (the differential time delay) is of more observational importance, we have shown it instead of explicit results for the secondary images. 
\begingroup
\begin{table*}
	\caption{
		{\bf Magnifications and time delays of primary and secondary images due to lensing by Sgr A* with ${\cal D}=0.5$}: GR and EQG predictions for magnifications $\mu$, time delays $\tau$, and differential time delays $t_d=\tau_s -\tau_p$ are given for different values of angular source position $\beta$. {\bf (a)} As in Table \ref{tab:Tablei}. {\bf (b)} $\beta$ is in {\em arcseconds} and time delays are in {\em minutes}. {\bf (c)} As in Table \ref{tab:Tablei}.
	}\label{tab:Tableii}
	\begin{tabular}{l cccc cccc}
		\hline \hline
		\multicolumn{1}{c}{$\beta$}&
		\multicolumn{4}{c}{General relativity}&
		\multicolumn{4}{c}{Einsteinian Quartic Gravity}\\
		&$\mu_{p,{\rm GR}} $&$\tau_{p,{\rm GR}}$&$\mu_{s,{\rm GR}}$&$t_{d,{\rm GR}}$&$\mu_{p,{\rm EQG}}$&$\tau_{p,{\rm EQG}}$&$\mu_{s,{\rm EQG}}$&$t_{d,{\rm EQG}}$\\
		\hline
		$0	 	 $&$	\times    $&$	16.588180 $&$ \times      $&$   0            $&$	\times    $&$	16.588781 $&$ \times      $&$   0   	        $\\
		$10^{-3} $&$	722.117   $&$	16.587267 $&$ -721.117    $&$	0.001830     $&$	721.788   $&$	16.586162 $&$ -720.789    $&$	0.003536        $\\
		$10^{-2} $&$	72.6630   $&$	16.579043 $&$ -71.6630    $&$	0.018298     $&$	72.6299   $&$	16.579647 $&$ -71.6303    $&$	0.018292        $\\
		$10^{-1} $&$	7.72915   $&$	16.498254 $&$ -6.72916    $&$	0.183013     $&$	7.72566   $&$	16.496903 $&$ -6.72607    $&$	0.184939      $\\
		$1	 	 $&$	1.34553   $&$	15.813792 $&$ -0.345536   $&$	1.865731     $&$	1.34500   $&$	15.814721 $&$ -0.345374   $&$	1.865202          $\\
		$2	 	 $&$	1.08134   $&$	15.254987 $&$ -0.0813405  $&$	3.934199     $&$	1.08099   $&$	15.256347 $&$ -0.0813025  $&$	3.933148         $\\
		$3	 	 $&$	1.02708   $&$	14.835066 $&$ -0.0270804  $&$	6.358812     $&$	1.02682   $&$	14.836917 $&$ -0.0270675  $&$	6.356546         $\\
		$4	 	 $&$ 	1.01102   $&$ 	14.505298 $&$ -0.0110231  $&$ 	9.237035     $&$ 	1.01085   $&$ 	14.500207 $&$ -0.0110166  $&$ 	9.242156        $\\
		\hline \hline
	\end{tabular}
\end{table*}
\endgroup

\begin{figure}[htp]
	\centering
	\includegraphics[width=1\textwidth]{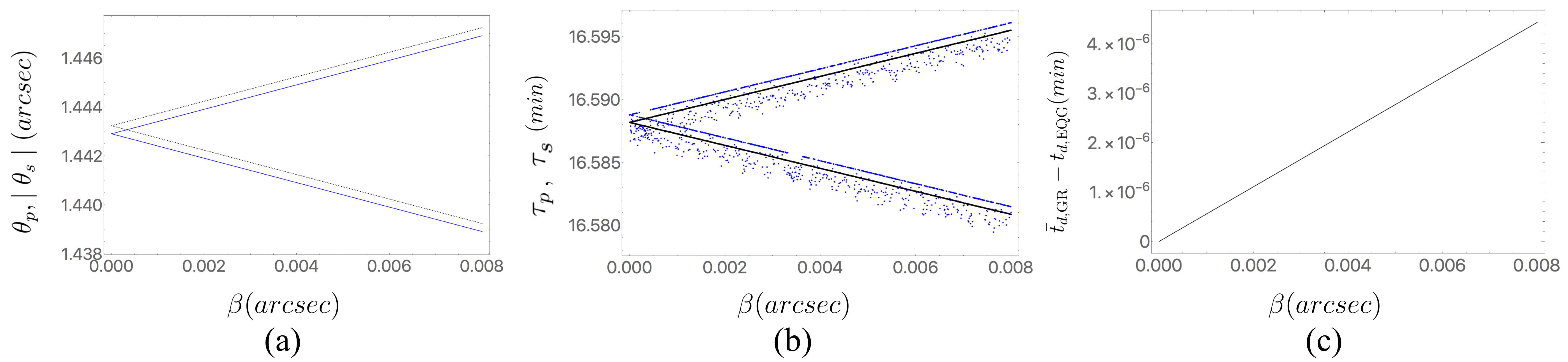}
	\caption{\textbf{Finding the source position}: \textbf{a}: Image positions as a function of the angular source position $\beta$ in GR (dotted, gray curve) and EQG (solid, blue curve) with $\mathcal{D}=0.5$.  Those lines with positive slope correspond to the primary image position $\theta_{p}$ and those with negative slope to the secondary image position $|\theta_{s}|$.
	\textbf{b}: The time delay in EQG with $\mathcal{D}=0.5$ (blue points) and in GR with $\mathcal{D}=0.499995$ (black points) as a function of the angular source position. It is easy to find a certain behavior for different value of $\beta$ in EQG like GR and the imaginary lines which passe through these points with positive slope correspond to the primary time delay $\tau_p$ and those with negative slope to the secondary time delay $\tau_s$.
	\textbf{c}: Difference between the differential time delay in GR with $\mathcal{D}=0.499995$, $\bar{t}_{d,{\rm GR}}$, and that in EQG with $\mathcal{D}=0.5$, $t_{d,{\rm ECG}}$ for a lots of number of $\beta$. We have used Sgr A* as the lens with $M_{Sgr A*}=5.94\times 10^9 \, {\rm m}$ and $D_d=2.43\times 10^{20} \, {\rm m}$, and have taken $K/M_{Sgr A*}^{6}\approx 2.21\times 10^{-1}$.
	}
	\label{fig:finding_source}
\end{figure} 

\begin{figure}
	\centering
	\includegraphics[width=1\textwidth]{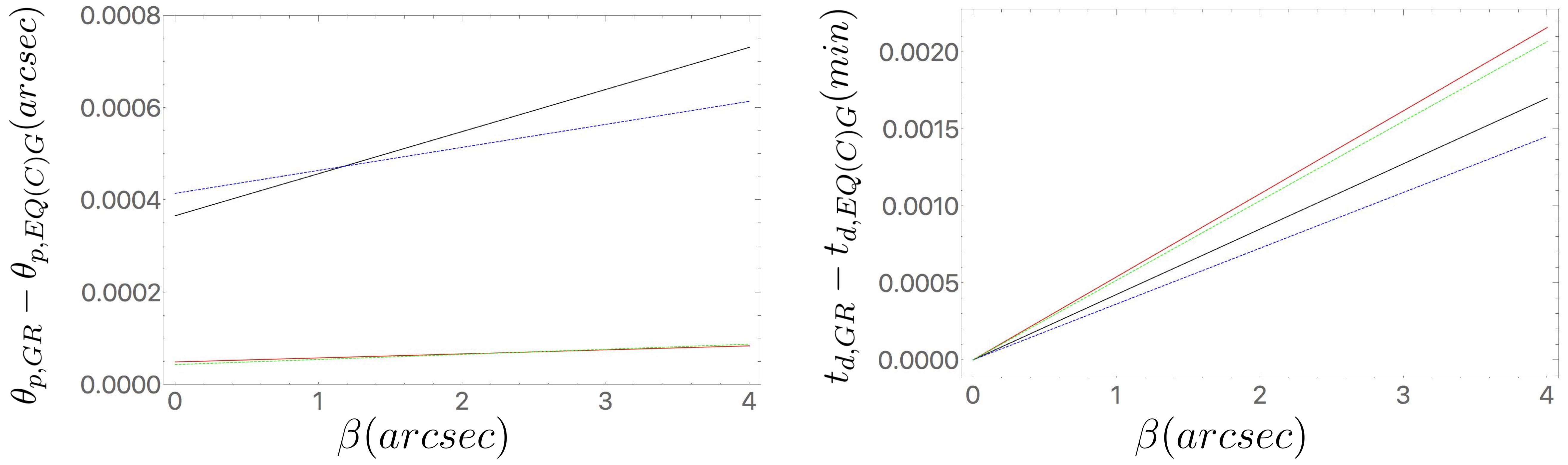}
	\caption{{\bf Deviation of primary image angular position and differential time delay in EQG and ECG 	
	from GR for Sgr A*}. The solid black 	
	(blue dotted) lines correspond to ${\cal D}=0.5$ and the red (green dotted) lines  to ${\cal D}=0.05$ in EQG (ECG). \textbf{Left:} Deviation plotted against angular source position $\beta$.  It is obvious that for a fixed lens-observer distance, the deviation of EQG (ECG)   for angular positions of primary images relative to that of GR is larger for sources further away from the lens. \textbf{Right}: Differential time delay $t_d=\tau_s-\tau_p$ 
plotted against angular source position $\beta$. We see that 	$t_d$  deviates from its corresponding value in GR if EQG (ECG) governs the strong gravitational field around the black hole. The deviation increases with increasing  angular source position $\beta$.  
Note that for small ${\cal D}$ the distinction between EQG and ECG is very tiny.
	}
	\label{fig:diff_theta}
\end{figure}

Suppose that the source is pulsating. Every phase in its period would then appear in the secondary image, $t_d$ minutes after it appears in the primary image. Comparing the results of GR and EQG in Table \ref{tab:Tableii}, it is obvious that for a wide range of $\beta$ the differential time delay $t_d$ is lower and EQG describes the strong gravitational field near the black hole. EQG, in addition, decrease the magnifications $\mu_p$ and $|\mu_s|$ by a small 
amount for larger  $\beta$.

Of course observationally it is the images that are detected and not the source itself. Although finding the distance $D_{ds}$ to the source from its redshift 
is possible under certain circumstances \cite{Falco}, the angular position $\beta$ is not directly observable. We can, however, adapt a scheme developed for
ECG to find $\beta$ from   primary and secondary image positions, time delays and their differential time delays to EQG.
In Fig.~\ref{fig:finding_source}a we   plot $\theta_{p}$ and $|\theta_{s}|$, the respective primary and secondary angular image positions 
in GR and EQG for $\mathcal{D}=0.5$. Each of these lines crosses both the plot of GR and EQG. We do not know if the theory governing the strong gravitational field is GR or EQG (assuming that one or the other is the empirically correct theory). However the correct theory must (for a given set of parameters) have the same value of $\beta$ at both intersection points, allowing for its determination.

In certain situations the distance to the source (and hence the value of $\mathcal{D}$) may not be known.  For example GR with $\mathcal{D}=0.499995$ yields almost the same lines for the image positions as EQG with $\mathcal{D}=0.5$ (the solid blue curves in Fig.~\ref{fig:finding_source}a). In other words, although $\beta$ can be distinguished via the intersection points of the $\theta_{p}$ and $|\theta_{s}|$ curves with observation, this is  insufficient to determine $\mathcal{D}$ and distinguish between GR and  EQG. In this case a measurement of the differential time delay could be used to break this degeneracy:  the time $\bar{t}_{d,{\rm GR}}$ it takes an image to reach an observer in GR is in general larger than that in EQG. Provided the images have sufficient temporal variability to measure the time $t_{obs}$ it takes the image to reach an observer, if the difference $\bar{t}_{d,{\rm GR}} - t_{obs}$ yields a value of $\beta$ consistent with	the aforementioned image observations, the value of  $\mathcal{D}$ could be inferred. We have shown two example in the Fig.~\ref{fig:finding_source}b and Fig.~\ref{fig:finding_source}c. In Fig.~\ref{fig:finding_source}b, we have plotted the primary and secondary time delay in EQG with $\mathcal{D}=0.5$ (blue points) and in GR with $\mathcal{D}=0.499995$ (black points). As we can see it is easy to find a certain behavior for different value of $\beta$ in EQG like GR. By considering these values of $\beta$ we have illustrated in Fig.~\ref{fig:finding_source}c that the differential time delay $t_d$ in GR with $\mathcal{D}=0.499995$ is larger than that in EQG with $\mathcal{D}=0.5$. In conjunction with an observation of the primary and secondary images, a time delay measurement can provide enough information to obtain $\beta$ and $\mathcal{D}$ and distinguish the governing theory of the gravitational field of the black hole.

GR and EQG results for magnifications, and the time delays of first and second order relativistic images are presented in Tables \ref{tab:Tableiii} and \ref{tab:Tableiv}, respectively. First (Second) order relativistic images are produced after the light winds, once (twice) around the black hole before reaching the observer \cite{Ellis}. Here noticeable differences with the corresponding results in ECG \cite{POSHTEH} are now apparent, with values of
$(\tau_{2p}-\tau_{1p})$ differing by as much as $30\%$ and of $\mu_{2p}$ by close to a factor of 2.
The angular position of relativistic images $\theta_{1p}$, $|\theta_{1s}|$, $\theta_{2p}$, and $|\theta_{2s}|$ are almost independent of angular source positions. In EQG their values are about $13$ and $4$ nanoarcseconds less than their corresponding values in GR for first and second order relativistic images respectively, an effect too tiny to be observed with today's telescopes, especially since these relativistic images are highly demagnified. 
However once technology develops the renders them observable,  (differential) time delays of relativistic images could be used to test both EQG and ECG, because of their increasing deviation from GR for large $\beta$, as can be seen from Tables \ref{tab:Tableiii} and \ref{tab:Tableiv}.

In Table \ref{tab:Table5} we present results for primary and secondary images in EQG when the source is closer to Sgr A*. In particular, we have taken ${\cal D}=0.05$. Comparing these results those in  Table \ref{tab:Tablei} (in which ${\cal D}=0.5$), shows that when the source-lens distance is smaller, primary and secondary images get closer to the line of sight to the lens ($\theta_{p}$ and $|\theta_{s}|$ get smaller).  Furthermore, a comparison of Tables \ref{tab:Table5} and \ref{tab:Tableii} shows that the magnification $\mu_{p}$ and $|\mu_{s}|$ and the time delay of the primary image are smaller in the case of ${\cal D}=0.05$ compared to ${\cal D}=0.5$. However the differential time delay $t_d=\tau_s-\tau_p$ is larger in the former case. Similar results hold when the governing theory of gravity is GR~\cite{Virbhadra}.

 In Fig.~\ref{fig:diff_theta} we illustrate some  relevant comparisons between GR, ECG and EQG. We see that the difference between  these three theories in   the angular positions of primary images is negligible for ${\cal D}=0.005$, but become distinguishable by parts in $10^{-4}$ -- $10^{-3}$  for ${\cal D}=0.5$.  
 Differences in the time delays for both ECG and EQG compared to GR become apparent by parts in $\sim 10^{-3}$ for large enough $\beta$, but the distinction
 between ECG and EQG are at least an order of magnitude smaller.
 
In Fig.~\ref{fig:diff_thetaeq} we directly compare  the results of EQG and ECG for  large values of their respective coupling constants.   We find that the difference between the results of EQG and ECG for the angular position of primary images is larger for the source further away from lens and is quite small for $\beta \approx 1$. The deviation of the differential time delay $t_d$ in EQG from its corresponding ECG is larger for the source further away from lens and increases by increasing $\beta$. Overall the distinctions are very small, not larger than $\sim 10^{-4}$, making it a formidable challenge to distinguish  the two theories from each other. using Sgr A*.

\begin{figure}
	\centering
	\includegraphics[width=0.5\textwidth]{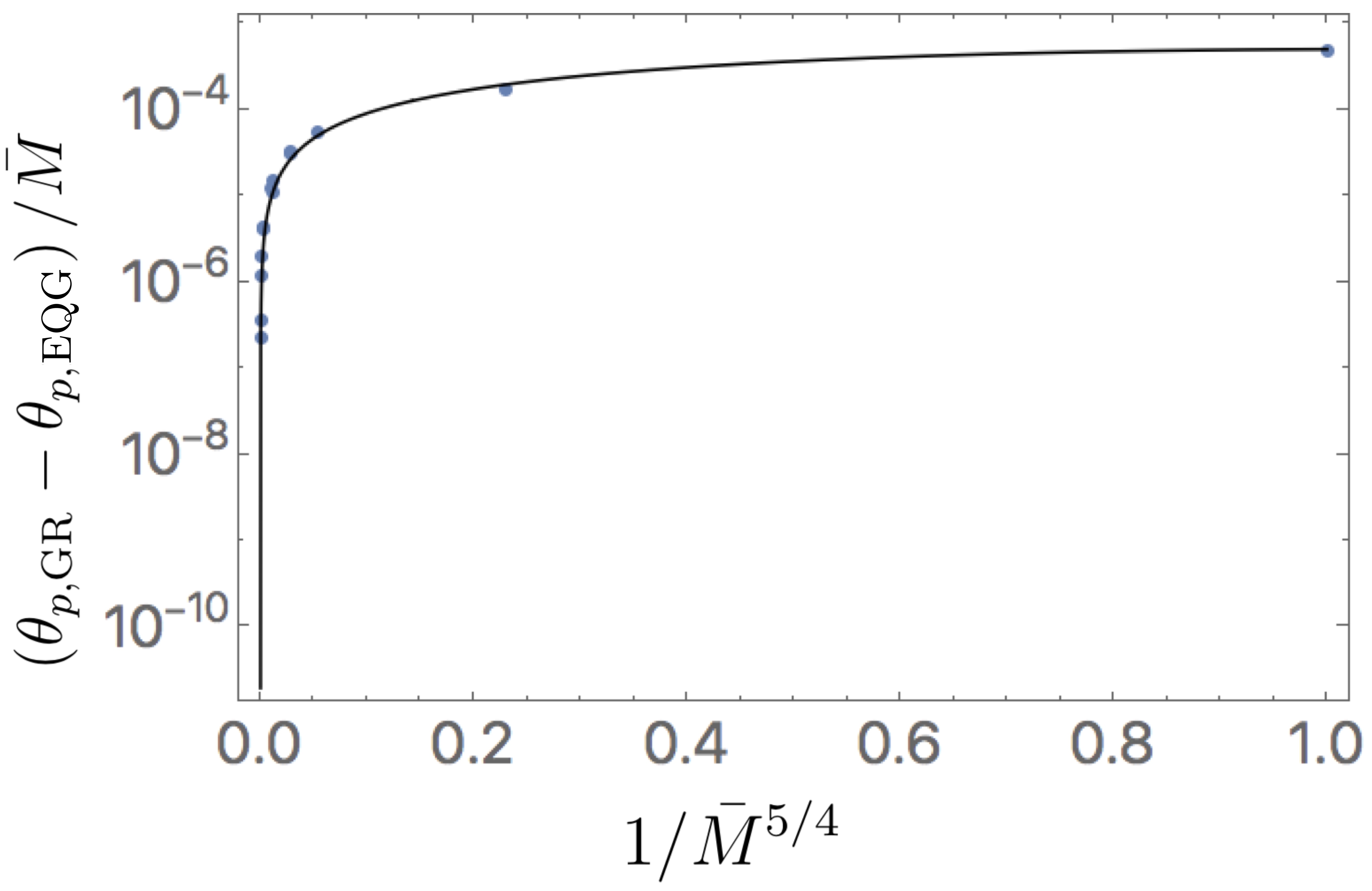}
	\caption{{\bf Deviation of primary image angular position in EQG from GR for different SMBHs}: The ratio $\left(\theta_{p,{\rm GR}}-\theta_{p,{\rm EQG}}\right)/\bar{M}$ increases as $\bar{M}$ decreases.  
	Here $\bar{M}=M/M_{Sgr A*}$, where $M$ is the mass of the SMBH from Table~\ref{tab:Table6}. We have taken ${\cal D}=0.5$.  The dots refer to the numerical results of Table \ref{tab:Table7} for the 14 SMBHs, and  the solid curve is the interpolation between the points.}
	\label{fig:angularposition}
\end{figure}

We close this section by considering SMBHs in other galaxies, whose   masses and distances  differ considerably from that of  Sgr A*. We collect in Table \ref{tab:Table6} some updated data of 14 galaxies  \cite{mnd,kormendy2013}, and  use this in Table \ref{tab:Table7} to calculate the time delays and angular positions of primary images in GR and EQG, as well as  between  secondary and primary images. 
Fig.~\ref{fig:angularposition} illustrates how the difference in the angular position of the primary image between GR and EQG changes with the mass of the black hole.  As previously mentioned   for Sgr A*,   differential time delays in EQG are quite sensitive to the angular source position $\beta$ and we cannot compare them for a special case $\beta=1$. The results Table \ref{tab:Table7} and Fig.~\ref{fig:angularposition} for EQG are quite close to the results for ECG \cite{POSHTEH}. Making it quite difficult to use  other galaxies to probe  empirical differences between EQG and ECG.

\begin{center}
	\begingroup
	\begin{table*}
		\caption{
			{\bf Magnifications and time delays of first order relativistic images due to lensing by Sgr A* with ${\cal D}=0.5$}: GR and EQG predictions for magnifications $\mu$ and time delays $\tau$ are given for different values of angular source position $\beta$. {\bf (a)} $1p$ and $1s$ refer to first order relativistic images on the same side as primary and secondary images, respectively. {\bf (b)} As in Table \ref{tab:Tableii}. {\bf (c)} As in Table \ref{tab:Tablei}. {\bf (d)} Angular positions of first order relativistic images in GR and EQG are, respectively, $\theta_{1p,{\rm GR}}\approx -\theta_{1s,{\rm GR}}\approx 26.2691 \mu as$ and $\theta_{1p,{\rm EQG}}\approx -\theta_{1s,{\rm EQG}}\approx 26.2560 \mu as$ and are highly insensitive to the angular source position $\beta$.
		}\label{tab:Tableiii}
		\begin{tabular}{l cccc cccc}
			\hline \hline
			\multicolumn{1}{c}{$\beta$}&
			\multicolumn{4}{c}{General relativity}&
			\multicolumn{4}{c}{Einsteinian Quartic Gravity}\\
			&$\mu_{1p,{\rm GR}} $&$\tau_{1p,{\rm GR}}$&$\mu_{1s,{\rm GR}}$&$\tau_{1s,{\rm GR}}$&$\mu_{1p,{\rm EQG}}$&$\tau_{1p,{\rm EQG}}$&$\mu_{1s,{\rm EQG}}$&$\tau_{1s,{\rm EQG}}$\\
			\hline
			$0        $&$	 \times	 		 $&$ 42.673253 $&$	   \times 		   $&$ 42.673253 $&$   \times		    $&$ 42.972656 $&$	  \times		  $&$ 42.972656         $\\
			$10^{-6}  $&$8.42\times 10^{-12} $&$ 42.673253 $&$-8.42\times 10^{-12} $&$ 42.673253 $&$6.33\times 10^{-12} $&$ 42.972656 $&$-6.33\times 10^{-12} $&$ 42.972656  		$\\
			$10^{-5}  $&$8.42\times 10^{-13} $&$ 42.673253 $&$-8.42\times 10^{-13} $&$ 42.673253 $&$6.33\times 10^{-13} $&$ 42.972656 $&$-6.33\times 10^{-13} $&$ 42.972656  		$\\
			$10^{-4}  $&$8.42\times 10^{-14} $&$ 42.673253 $&$-8.42\times 10^{-14} $&$ 42.673253 $&$6.33\times 10^{-14} $&$ 42.972656 $&$-6.33\times 10^{-14} $&$ 42.972656  		$\\
			$10^{-3}  $&$8.42\times 10^{-15} $&$ 42.673255 $&$-8.42\times 10^{-15} $&$ 42.673255 $&$6.33\times 10^{-15} $&$ 42.972659 $&$-6.33\times 10^{-15} $&$ 42.972659 		$\\
			$10^{-2}  $&$8.42\times 10^{-16} $&$ 42.673280 $&$-8.42\times 10^{-16} $&$ 42.673281 $&$6.33\times 10^{-16} $&$ 42.972684 $&$-6.33\times 10^{-16} $&$ 42.972684  		$\\
			$10^{-1}  $&$8.42\times 10^{-17} $&$ 42.676417 $&$-8.42\times 10^{-17} $&$ 42.676420 $&$6.33\times 10^{-17} $&$ 42.975822 $&$-6.33\times 10^{-17} $&$ 42.975822  		$\\
			$1        $&$8.42\times 10^{-18} $&$ 42.990190 $&$-8.42\times 10^{-18} $&$ 42.990223 $&$6.33\times 10^{-18} $&$ 43.289610 $&$-6.33\times 10^{-18} $&$ 42.289610  		$\\
			\hline \hline
		\end{tabular}
	\end{table*}
	\endgroup
\end{center}

\begingroup
\begin{table*}
	\caption{
		{\bf Magnifications and time delays of second order relativistic images due to lensing by Sgr A* with ${\cal D}=0.5$}: GR and EQG predictions for magnifications $\mu$, time delays $\tau$, and differential time delays $\tau_{2p}-\tau_{1p}$ are given for different values of angular source position $\beta$. {\bf (a)} $2p$ and $2s$ refer to second order relativistic images on the same side as primary and secondary images, respectively. {\bf (b)} As in Table \ref{tab:Tableii}. {\bf (c)} As in Table \ref{tab:Tablei}. {\bf (d)} Angular positions of second order relativistic images in GR and EQG are, respectively, $\theta_{2p,{\rm GR}}\approx -\theta_{2s,{\rm GR}}\approx 26.2362 \mu as$ and $\theta_{2p,{\rm EQG}}\approx -\theta_{2s,{\rm EQG}}\approx 26.2313 \mu as$ and are highly insensitive to the angular source position $\beta$. {\bf (e)} $\mu_{2s}= -\mu_{2p}$ to a very good approximation. {\bf (f)} Explicit values of $\tau_{1p}$ are given in Table \ref{tab:Tableiii}.
	}\label{tab:Tableiv}
	\begin{tabular}{l cccc cccc}
		\hline \hline
		\multicolumn{1}{c}{$\beta$}&
		\multicolumn{4}{c}{General relativity}&
		\multicolumn{4}{c}{Einsteinian Quartic Gravity}\\
		&$\mu_{2p,{\rm GR}}$&$\tau_{2p,{\rm GR}}$&$\tau_{2s,{\rm GR}}$&$(\tau_{2p}-\tau_{1p})_{{\rm GR}}$&$\mu_{2p,{\rm EQG}}$&$\tau_{2p,{\rm EQG}}$&$\tau_{2s,{\rm EQG}}$&$(\tau_{2p}-\tau_{1p})_{{\rm EQG}}$\\
		\hline
		$0        $&$     \times              $&$ 53.452468 $&$ 53.452468 $&$ 10.779215$&$     \times              $&$ 50.146865 $&$ 50.146865 $&$ 7.174209       $\\
		$10^{-6}  $&$     1.56\times 10^{-14} $&$ 53.452468 $&$ 53.452468 $&$ 10.779215$&$     4.33\times 10^{-14} $&$ 50.146865 $&$ 50.146865 $&$ 7.174209  		$\\
		$10^{-5}  $&$     1.56\times 10^{-15} $&$ 53.452468 $&$ 53.452468 $&$ 10.779215$&$     4.33\times 10^{-15} $&$ 50.146865 $&$ 50.146865 $&$ 7.174209  		$\\
		$10^{-4}  $&$     1.56\times 10^{-16} $&$ 53.452468 $&$ 53.452468 $&$ 10.779215$&$     4.33\times 10^{-16} $&$ 50.146865 $&$ 50.146865 $&$ 7.174209  		$\\
		$10^{-3}  $&$     1.56\times 10^{-17} $&$ 53.452471 $&$ 53.452471 $&$ 10.779215$&$     4.33\times 10^{-17} $&$ 50.146867 $&$ 50.146867 $&$ 7.174209 		$\\
		$10^{-2}  $&$     1.56\times 10^{-18} $&$ 53.452496 $&$ 53.452496 $&$ 10.779215$&$     4.33\times 10^{-18} $&$ 50.146892 $&$ 50.146892 $&$ 7.174208  		$\\
		$10^{-1}  $&$     1.56\times 10^{-19} $&$ 53.455632 $&$ 53.455635 $&$ 10.779215$&$     4.33\times 10^{-19} $&$ 50.150031 $&$ 50.150031 $&$ 7.174208  		$\\
		$1        $&$     1.56\times 10^{-20} $&$ 53.769405 $&$ 53.769438 $&$ 10.779215$&$     4.33\times 10^{-20} $&$ 50.463818 $&$ 50.463818 $&$ 7.174208  		$\\
		\hline \hline
	\end{tabular}
\end{table*}
\endgroup

\begingroup
\begin{table*}
	\caption{
		{\bf Primary and secondary images due to lensing by Sgr A* in EQG with ${\cal D}=0.05$}: Angular  positions $\theta$, bending angles $\hat{\alpha}$, magnifications $\mu$, time delays $\tau$, and the differential time delay $t_d=\tau_s-\tau_p$ are given for different values of angular source position $\beta$. {\bf (a)} As in Table \ref{tab:Tablei}. {\bf (b)} All angles are in {\em arcseconds} and time delays are in {\em minutes}. {\bf (c)} As in Table \ref{tab:Tablei}.
	}\label{tab:Table5}
	\begin{tabular}{l cccc cccc}  
		\hline \hline
		$\beta	 $&$  \theta_p    $&$\hat{\alpha}_p$&$   \mu_p    $&$  \tau_p     $&$ \theta_{s}$&$\hat{\alpha}_{s}$&$  \mu_{s}    $&$  t_d			   $\\
		\hline
		$0	 	 $&$  0.45635     $&$	9.12842    $&$	\times    $&$	16.164730 $&$ -0.45635  $&$    9.12842     $&$ \times      $&$  0    		   $\\
		$10^{-3} $&$  0.45685	  $&$   9.11744    $&$	228.650   $&$	16.161839 $&$ -0.45585  $&$	   9.13897     $&$ -227.650    $&$	0.005786       $\\
		$10^{-2} $&$  0.46138	  $&$   9.02670    $&$	23.3190   $&$	16.135598 $&$ -0.45138  $&$	   9.22620     $&$ -22.3191    $&$	0.057753       $\\
		$10^{-1} $&$  0.50908	  $&$   8.18338    $&$	2.82231   $&$	15.890705 $&$ -0.40908  $&$	   10.1821     $&$ -1.82244    $&$	0.579745       $\\
		$1	 	 $&$  1.17691	  $&$   3.53939    $&$	1.02306   $&$	14.353080 $&$ -0.17697  $&$	   23.5392     $&$ -0.02313    $&$	6.792761       $\\
		$2	 	 $&$  2.09916	  $&$   1.98512    $&$	1.00220   $&$	13.521525 $&$ -0.09923  $&$    41.9845     $&$ -0.00224    $&$	17.96517       $\\
		$3	 	 $&$  3.06782	  $&$   1.35540    $&$	1.00046   $&$	13.002012 $&$ -0.06791  $&$	   61.3628     $&$ -0.00049    $&$	34.84999       $\\
		$4	 	 $&$  4.05133	  $&$   1.02451    $&$ 	1.00014   $&$ 	12.626937 $&$ -0.05143  $&$    81.0278     $&$ -0.00016    $&$ 	57.78525       $\\
		\hline \hline
	\end{tabular}
\end{table*}
\endgroup

\begingroup
\begin{table*} 
	\caption{ 
		{\bf Masses and distances of SMBHs}: Masses ($M$) and distances ($D_d$) of SMBHs at the center of 14 galaxies. The data for Sgr A* at the center of Milky Way Galaxy has been taken from~\cite{mnd}. The data of other black holes are from~\cite{kormendy2013}.
	}\label{tab:Table6}
		\begin{tabular}{cccc cccc}  
			\hline \hline
			Galaxy   &           $M$ (m)      &          $D_d$ (m)   &     $D_d/M$        & Galaxy   &         $M$ (m)      &       $D_d$ (m)      &      $D_d/M$         \\
			\hline
			Milky Way&$  5.94\times 10^9	 $&$ 2.43\times 10^{20} $&$4.09\times 10^{10}$& M31      &$ 2.11\times 10^{11} $&$ 2.39\times 10^{22} $&$1.13\times 10^{11}$     \\
			M87      &$  9.08\times 10^{12}  $&$ 5.15\times 10^{23} $&$5.67\times 10^{10}$& NGC 1023 &$ 6.10\times 10^{10} $&$ 3.34\times 10^{23} $&$5.48\times 10^{12}$      \\
			NGC 1194 &$  1.05\times 10^{11}  $&$ 1.79\times 10^{24} $&$1.70\times 10^{13}$& NGC 1316 &$ 2.50\times 10^{11} $&$ 6.47\times 10^{23} $&$2.59\times 10^{12}$      \\
			NGC 1332 &$  2.17\times 10^{12}  $&$ 6.99\times 10^{23} $&$3.22\times 10^{11}$& NGC 1407 &$ 6.87\times 10^{12} $&$ 8.95\times 10^{23} $&$1.30\times 10^{11}$     \\
			NGC 3607 &$  2.02\times 10^{11}  $&$ 6.99\times 10^{23} $&$3.46\times 10^{12}$& NGC 3608 &$ 6.87\times 10^{11} $&$ 7.02\times 10^{23} $&$1.02\times 10^{12}$    \\
			NGC 4261 &$  7.81\times 10^{11}  $&$ 9.99\times 10^{23} $&$1.28\times 10^{12}$& NGC 4374 &$ 1.37\times 10^{12} $&$ 5.71\times 10^{23} $&$4.17\times 10^{11}$      \\
			NGC 4382 &$  1.92\times 10^{10}  $&$ 5.52\times 10^{23} $&$2.88\times 10^{13}$& NGC 4459 &$ 1.03\times 10^{11} $&$ 4.94\times 10^{23} $&$4.80\times 10^{12}$      \\
			\hline \hline
		\end{tabular}
\end{table*}
\endgroup

\begingroup
\begin{table*}
	\caption{
		{\bf Image positions and time delays due to lensing by SMBHs}: GR and EQG predictions for angular positions $\theta$ and the time delays $\tau$ of primary images as well as the differential time delays $t_d=\tau_s-\tau_p$ are given for different SMBHs. We have also presented the difference between GR and EQG predictions $\theta_p$ and $t_d$. {\bf (a)} As in Table \ref{tab:Tablei}. {\bf (b)} All angles are in {\em arcseconds} and time delays are in {\em minutes}. {\bf (c)} We have taken ${\cal D}=0.5$, $\beta =1 arcsecond$, and $K \approx 9.32 \times 10^{57}$.
	}\label{tab:Table7}
		\begin{tabular}{l ccc ccccc}
			\hline \hline
			\multicolumn{1}{c}{Galaxy}&
			\multicolumn{3}{c}{General relativity}&
			\multicolumn{5}{c}{Einsteinian Cubic Gravity}\\
			&$\theta_{p,{\rm GR}}$&$  \tau_{p,{\rm GR}}  $&$  t_{d,{\rm GR}}  $&$ \theta_{p,{\rm EQG}} $&$ \tau_{p,{\rm EQG}} $&$  t_{d,{\rm EQG}}  $&$\theta_{p,{\rm GR}}-\theta_{p,{\rm EQG}}$&$t_{d,{\rm GR}}-t_{d,{\rm EQG}}$\\
			\hline
			Milky Way &$  2.02740    $&$   15.813781   $&$  1.865927  $&$  	 2.05382  $&$  15.81.4721  $&$  1.865202 $&$ 0.00048 $&$ 0.000529 $\\
			M31       &$  1.50121    $&$   572.58323   $&$  114.0235  $&$  	 1.50082  $&$  572.54559  $&$ 114.0786 $&$ 0.00039 $&$ -0.055106 $\\
			M87  &$  1.82348    $&$   24351.162   $&$  3386.221  $&$  	 1.82312  $&$  24348.702  $&$ 3389.338 $&$  0.00035  $&$ -3.117026 $\\
			NGC 1023  &$  1.01532    $&$   168.51053   $&$  506.3728  $&$  	 1.01477  $&$ 167.96251 $&$ 506.9039 $&$ 0.00055 $&$ -0.531172 $\\
			NGC 1194  &$  1.00495    $&$   288.99871   $&$  2485.346  $&$  	 1.00442   $&$ 291.49859 $&$ 2482.817  $&$ 0.00053 $&$ 2.528745 $\\
			NGC 1316  &$  1.03184    $&$   689.09579   $&$  1091.665  $&$  	 1.03134   $&$ 688.38118  $&$ 1092.422 $&$ 0.00050 $&$ -0.756943 $\\
			NGC 1332  &$  1.21708    $&$   5949.7675   $&$  2142.847  $&$  	 1.21665   $&$  5949.6420 $&$ 2143.185 $&$ 0.00043  $&$ -0.337623  $\\
			NGC 1407  &$  1.45027    $&$   18653.830   $&$  4008.966  $&$  	 1.44985    $&$  18655.717  $&$ 4006.356 $&$ 0.00042 $&$ 2.610108 $\\
			NGC 3607  &$  1.02405    $&$   558.79217   $&$  1126.096  $&$  	 1.02354   $&$  559.78355 $&$  1125.099  $&$ 0.00052 $&$ 0.996905 $\\
			NGC 3608  &$  1.07727    $&$   1892.5635   $&$  1461.592  $&$  	 1.07678     $&$ 1893.6109  $&$ 1460.623 $&$ 0.00049 $&$ 0.968664 $\\
			NGC 4261  &$  1.06265    $&$   2154.3003   $&$  1960.302  $&$  	 1.06212    $&$ 2152.8429  $&$ 1961.835 $&$ 0.00054  $&$ -1.532530 $\\
			NGC 4374  &$  1.17344    $&$   3750.3521   $&$  1586.191  $&$  	 1.17299    $&$ 3748.7532 $&$ 1587.936 $&$ 0.00045 $&$ -1.744682 $\\
			NGC 4382  &$  1.00295    $&$   53.070221   $&$  750.1529  $&$  	 1.00239  $&$  52.941942  $&$  750.2743 $&$ 0.00056 $&$ -0.121390 $\\
			NGC 4459  &$  1.01740    $&$   283.95236   $&$  761.0761  $&$  	 1.01685  $&$  283.82393 $&$  761.2151 $&$ 0.00055 $&$ -0.139078 $\\
			\hline \hline
		\end{tabular}
\end{table*}
\endgroup

\section{Concluding Remarks} 

We have investigated GR and EQG prediction for GL effects by some SMBHs in our galaxy and other thirteen galaxies. By considering the Sgr A* as the lens and EQG coupling constant as $K = 8.98 \times 10^{38} M_{\astrosun}^4$, for which EQG passes all Solar System tests to date~\cite{Khodabakhshi:2020hny}, we calculated the angular positions of primary and secondary images deviation from that of GR by an amount of order of miliarcseconds. As well, using numerical methods we have generally shown that the EQG results for the differential time delay, associated with primary and secondary images, could be some tenths of seconds shorter than the results of GR for a lots of number of $\beta$ (please see Fig.~\ref{fig:finding_source}). 

It is important to note that for the primary/secondary images to be produced, the light from the source should pass the black hole at a closest distance of order $10^{5}\, r_+$, where $r_+$ is the radius of event horizon. We have illustrated even in this large distance from the black hole that EQG effects may be observable. One does not have to observe gravitational effects in the vicinity of an horizon to test EQG.

There are several short period stars (the so-called S-stars) orbiting around Sgr A* whose semimajor axes are less than $10^{5}\, r_+$ \cite{gillessen2017}. Nowadays the observation of these S-stars are possible with good precision \cite{abuter2017}. We propose, as a direction of future study, to investigate the orbit of S-stars in EQG and to compare it with observational results now available \cite{abuter2017,refId0}.

As for GR~\cite{Ellis,Virbhadra} and ECG \cite{POSHTEH}, in EQG relativistic images are produced after the light winds around the black hole. For these images to be produced the light must pass the black hole very closely. Consider the first order relativistic image. The closest approach of the light is $\sim 1.55 \, r_+$, which is very close to the radius of the photon sphere, $r_{ps}=1.5 \, r_+$, where the shadow is produced. The light must get closer and closer to the photon sphere to produce higher and higher order relativistic images.

We have shown in our previous paper~\cite{Khodabakhshi:2020hny} that the effects of EQG on the angular radius of the shadow of Sgr A* is less than 10 nanoarcseconds. Here we see that the same thing is also true for the angular positions of relativistic images. In this case the differential time delay between relativistic images could be used to test EQG, if (since they are highly demagnified) these images could ever be observed. 

We also considered GR and EQG predictions for lensing effects due to SMBHs in other galaxies. We find that results of GR and EQG for the differential time delay between primary and secondary images could differ by an amount of more than one minute for distant SMBHs. The deviation between GR and EQG predictions for image angular positions mostly depends on the mass of black hole and it reminds us to what we found in \cite{Khodabakhshi:2020hny}. Very massive EQG black holes are like ordinary Schwarzschild black holes. However intermediate mass EQG black holes deviate significantly. This point should be discussed elsewhere. 

Although our results provide some cautious optimism for distinguishing EQG from GR by observation,  tables \ref{tab:Tablei} and \ref{tab:Tableii}
indicate that using Sgr A* to distinguish  EQG from corresponding predictions in   ECG \cite{POSHTEH} will be very challenging. In Figs.~\ref{fig:diff_theta} and ~\ref{fig:diff_thetaeq} we compared deviation of the primary image angular position and differential time delay in EQG from ECG for Sgr A*. The deviation of EQG results for angular positions of primary images relative to that in ECG is larger for  sources that are further away from the lens, but quite small for $\beta \approx 1$. Furthermore, the deviation of EQG results for the differential time delay $t_d=\tau_s-\tau_p$ relative to  those in ECG is larger for the sources further away from lens and increase with  increasing $\beta$. The results for the relativistic images in Tables \ref{tab:Tableiii} and \ref{tab:Tableiv} show that EQG results are small but they have noticeable differences with ECG results \cite{POSHTEH}.  The situation is not much better for black holes in other galaxies;  table~\ref{tab:Table7} 
indicates that EQG predictions are quite similar to those in ECG \cite{POSHTEH} for the largest possible values of their respective coupling constants. In fact for other galaxies like Sgr A*, EQG and ECG results are quite similar for the special choice ${\cal D}=0.5$ and $\beta \approx1$ (see Fig.~\ref{fig:diff_thetaeq}).  
While we might hope to distinguish (or place bounds on) non-linear curvature effects of GQTGs using gravitational lensing, it will be very challenging to distinguish between the two simplest GQTGs. One possibility is to incorporate rotation. Slowly-rotating solutions have been obtained 
numerically in ECG \cite{Adair:2020vso}, allowing for a computation of the photon sphere and the innermost stable circular orbit.  It is reasonable to expect that there will be distinct features that not only distinguish the ECG predictions from EQG, but also distinguish the different EQG theories from each other.

\section*{Acknowledgments}

This work was supported in part by the Natural Sciences and Engineering Research Council of Canada.

\end{document}